\documentclass[aps,prd,preprint,groupedaddress,showpacs]{revtex4-1}

\usepackage{amsmath,amssymb}
\usepackage[dvips]{graphicx}
\usepackage{braket}

\begin{document}


\preprint{OU-HET 738/2012}

\title{Entanglement Entropy of Two Spheres}



\author{Noburo Shiba}
\email{shiba@het.phys.sci.osaka-u.ac.jp }
\affiliation{Department of Physics, Graduate School of Science,
Osaka University, Toyonaka, Osaka 560-0043, Japan}


\date{\today}

\begin{abstract}
We study the entanglement entropy $S_{AB}$ of a massless free scalar field on two spheres $A$ and $B$ whose radii are $R_1$ and $R_2$, 
respectively, and the distance between the centers of them is $r$.
The state of the massless free scalar field is the vacuum state. 
We obtain the result that the mutual information $S_{A;B} \equiv  S_{A}+S_{B}-S_{AB}$ 
is independent of the ultraviolet cutoff and proportional to the product of the areas of the two spheres when $r\gg R_1,R_2$, 
where $S_A$ and $S_B$ are the entanglement entropy on the inside region of $A$ and $B$, respectively.
We discuss possible connections of this result 
with the physics of black holes.



\end{abstract}

\maketitle

\section{\label{sec level1}Introduction}

Entanglement entropy in the quantum field theory (QFT) was originally studied to explain black hole entropy \cite{Bombelli:1986rw,Srednicki:1993im}.
Entanglement entropy is generally defined as the von Neumann entropy $S_A=-Tr \rho_A \ln \rho_A$ corresponding to the reduced density matrix $\rho_A$ of a subsystem $A$.
When we consider the quantum field theory in $(d+1)$-dimensional spacetime $\mathbb{R}\times N$, where $\mathbb{R}$ and $N$ 
denote the time direction and the $d$-dimensional spacelike manifold, respectively,
we define the subsystem by a $d$-dimensional domain 
$A\subset N$ at fixed time $t=t_0$.
(So this is also called geometric entropy.)
Entanglement entropy naturally arises when we consider the black hole because we cannot 
obtain the information inside the black hole.
In fact,  in the vacuum state the leading term of the entanglement entropy of $A$ is proportional to the area of the boundary $\partial A$ in many cases 
\cite{Bombelli:1986rw,Srednicki:1993im}.
This is similar to black hole entropy, and extensive 
studies have been carried out \cite{Hawking:2000da, Kabat:1995eq, Susskind:1994sm, Frolov:1993ym, Jacobson:1994iw, 'tHooft:1984re}.

In this paper, we study the entanglement entropy $S_{AB}$ of the massless free scalar field in $(d+1)$-dimensional Minkowski spacetime on 
two spheres $A$ and $B$ whose radii are $R_1$ and $R_2$
and how it depends on the distance $r$ between the centers of the two spheres. 
Entanglement entropy of two disconnected regions has been studied, see e.g. \cite{Casini:2008wt,Calabrese:2009ez,Calabrese:2010he}.
We consider the case that the state of the massless free scalar field is the vacuum state. 
We studied $S_{AB}$ in \cite{Shiba:2010dy} analytically.
When $r\gg R_1,R_2$, we obtained the $r$ dependence of $S_{AB}$ as 
\begin{equation}
S_{AB}\approx S_A+S_B-\dfrac{G(R_1,R_2,a)}{r^{2d-2}}, \label{eq:1-1}
\end{equation}  
where $a$ is an ultraviolet cutoff length and $G(R_1,R_2,a)=G(R_2,R_1,a)\geq 0$.
(Notice that we defined $G(R_1,R_2,a)$ in (\ref{eq:1-1}) as that in \cite{Shiba:2010dy} multiplied by $(-1)$ for simplicity.)
We could not determine the functional form of $G(R_1,R_2,a)$. 
In this paper, we numerically calculate $S_{AB}$ for $d=2,3$. 
We obtain the result that the mutual information $S_{A;B} \equiv  S_{A}+S_{B}-S_{AB}$ 
is independent of the ultraviolet cutoff length and $G(R_1,R_2,a)$ is proportional to the simple product of the surface areas of two spheres. 
(Note that we cannot determine the functional form of $G(R_1,R_2)$ only by the constraints from dimensional analysis, symmetry, and behavior in the limit $R_1\rightarrow 0$.    
For example, $G(R_1,R_2)=R_1^3 R_2+R_1 R_2^3$ is not prohibited by these constraints.)   
The mutual information is a quantity that measures the entanglement between two systems. 
(See e.g. \cite{nielsen2000quantum})
In order to examine whether only the degrees of freedom on the surface of the spheres contribute to the mutual information or not, 
we calculate the mutual information $S_{D;E}$ of two same spherical shells $D$ and $E$ for $d=3$ and 
the mutual information $S_{H;I}$ of two same rings $H$ and $I$ for $d=2$. 
The internal (external) radii of the spherical shell and the ring are $L_1$ ($L_2$). 
The distance between the centers of the two spherical shells and that between the two rings are $r$. 
We obtain the result that $S_{D;E}$ and $S_{H;I}$ are monotone decreasing function of $L_1$.  
Then not only the degrees of freedom on the surface of the sphere but also those on the inside region contribute to the mutual information.  
This result is remarkably different from that of the entanglement entropy to which the degrees of freedom on the surface of the boundary contribute mainly. 



Previously, we studied $S_{AB}$ in \cite{Shiba:2010dy} in order to study
an entropic contribution to the force between two black holes.
To a distant observer, an object falling into a black hole takes
 an infinite time to reach the event horizon and the outside region is isolated from the inside region 
if we neglect the change of the mass of the black hole. 
Then we are probably able to consider the entanglement entropy of quantum fields on the outside region $C$ of two black holes $A$ and $B$ as thermodynamic entropy, 
and we can see the entropic force acting on the two black holes from the $r$ dependence of $S_C$. 
We consider two systems $X$ and $Y$, then one can show $S_X=S_Y$ in general if a composite system $XY$ is in a pure state.  
Then $S_C=S_{AB}$ when the state of the field on the whole space is a pure state. 
We will roughly estimate the magnitude of the entropic force between two black holes by using $S_{AB}$ in Minkowski spacetime.

The present paper is organized as follows. 
There have been some computational methods of entanglement entropy \cite{Calabrese:2004eu,Holzhey:1994we,Ryu:2006bv} 
and the reader is urged to refer to \cite{Ryu:2006ef,Casini:2009sr,Solodukhin:2011gn} for reviews. 
Among several others, we review in Sec.\ref{review} the method of Bombelli et al \cite{Bombelli:1986rw} 
which is most straightforward and suitable for numerical calculations.
In Sec.\ref{lattice}, 
we apply the above formalism to a massless free scalar field in $(d+1)$-dimensional Minkowski spacetime. 
We improve the computational method of Bombelli et al to reduce the computational complexity.  
In Sec.\ref{results}, we numerically calculate the entanglement entropy $S_{AB}$, the mutual information $S_{D;E}$, and $S_{H;I}$ 
in $(d+1)$-dimensional Minkowski spacetime for $d=2,3$.
We roughly estimate the magnitude of the entropic force between two black holes by $S_{AB}$ in (3+1)-dimensional Minkowski spacetime. 



\section{how to compute entanglement entropy} \label {review}
In this section we review the computational method developed by Bombelli et al \cite{Bombelli:1986rw}.
As a model amenable to unambiguous calculation we deal with the scalar field 
on $\mathbb{R}^d$ as a collection of coupled oscillators on a lattice of space points, 
labeled by capital Latin indices, the displacement at each point giving the value of the scalar field there.
In this case the Lagrangian can be given by
\begin{equation}
L=\dfrac{1}{2}G_{MN} \dot q ^M \dot q^N - \dfrac{1}{2} V_{MN} q^M q^N , \label{eq:3-1}
\end{equation}
where $q^M$ gives the displacement of the Mth oscillator and $\dot q^M$ its generalized velocity. 
The symmetric matrix $G_{MN}$ is positive definite and therefore invertible; i.e., there exists the inverse matrix $G^{MN}$ such that
\begin{equation}
G^{MP} G_{PN} = \delta ^M_{~~ N} . \label{eq:3-2}
\end{equation}
The matrix $V_{MN}$ is also symmetric and positive definite.
The matrices $G_{MN}$ and $V_{MN}$ are independent of $q^M$ and $\dot q^M$.
Introducing the conjugate momentum to $q^M$,
\begin{equation}
P_{M}=G_{MN} \dot q ^N , \label{eq:3-2-1}
\end{equation}
we can write the Hamiltonian for our system as
\begin{equation}
H=\dfrac{1}{2}G^{MN} P_M P_N + \dfrac{1}{2} V_{MN} q^M q^N . \label{eq:3-2-2}
\end{equation}
Next, consider the positive definite symmetric matrix $W_{MN}$ defined by
\begin{equation}
W_{MA} G^{AB} W_{BN} = V_{MN} . \label{eq:3-3}
\end{equation}
In this sense the matrix $W$ is the "square root" of $V$ in the scalar product with $G$.

Now consider a region $\Omega $ in $\mathbb{R}^d$.
The oscillators in this region will be specified by Greek letters, 
and those in the complement of $\Omega $, $\Omega ^c$, will be specified by lowercase Latin letters. 
We will use the following notation
\begin{alignat}{2}
W_{AB} = \begin{pmatrix} 
             W_{ab} & W_{a\beta}  \\
             W_{\alpha b} & W_{\alpha \beta} 
             \end{pmatrix} 
             \equiv   \begin{pmatrix} 
             A & B  \\
             B^T & C 
             \end{pmatrix}  &   ~~~~~~
 W^{AB} = \begin{pmatrix} 
             W^{ab} & W^{a\beta}  \\
             W^{\alpha b} & W^{\alpha \beta} \end{pmatrix}
              \equiv   \begin{pmatrix} 
             D & E  \\
             E^T & F 
             \end{pmatrix}             
, \label{eq:3-4}
\end{alignat}
where $W^{AB}$ is the inverse matrix of $W_{AB}$ ($W^{AB}$ is \textit{not} obtained by raising indices with $G^{AB}$).
So we have 
\begin{alignat}{2}
 \begin{pmatrix} 
             1 & 0  \\
             0 & 1 
             \end{pmatrix} 
             =   \begin{pmatrix} 
             A & B  \\
             B^T & C 
             \end{pmatrix}  
   \begin{pmatrix} 
             D & E  \\
             E^T & F 
             \end{pmatrix}  
             =     \begin{pmatrix} 
            A D+BE^T & AE+BF  \\
             B^T D+CE^T & B^TE+ CF 
             \end{pmatrix}        
. \label{eq:3-5}
\end{alignat}
  
If the information on the displacement of the oscillators in $\Omega $ is considered as unavailable, we can obtain a reduced density matrix 
$\rho_{red}$ for $\Omega^c $, integrating out over $q^{\alpha }\in \mathbb{R}$ for each of the oscillators in the region $\Omega $, and then we have
\begin{equation}
\rho_{red} ({q^a} , {q'^{b}}) =\int \prod_{\alpha } dq^{\alpha } \rho ({q^a, q^{\alpha }} , {q'^b,q^{\alpha}}) , \label{eq:3-6}
\end{equation}
where $\rho$ is a density matrix of the total system.

We can obtain the density matrix for the ground state by the standard method, and it is a Gaussian density matrix.
Then,  $\rho_{red}$ is obtained by a Gaussian integral, and it is also a Gaussian density matrix.
The entanglement entropy $S=-tr \rho_{red} \ln \rho_{red} $ is given by \cite{Bombelli:1986rw}
\begin{gather}
 S=     \sum_n f(\lambda_n) ,  \label{eq:new3-1}  \\  
   f(\lambda) \equiv   \ln (\dfrac{1}{2} \lambda ^{1/2} ) + (1+\lambda  )^{1/2 } \ln [(1+\lambda  ^{-1})^{1/2 } + \lambda ^{-1/2} ]   ,
     \label{eq:3-7}
\end{gather}
where $\lambda_n$ are the eigenvalues of the matrix 
\begin{equation}
\Lambda ^a _{~b} = - W^{a\alpha} W_{\alpha b} =-(E B^T)^a_{~~b} =(DA)^a_{~~b} -\delta^a_{~~b} .    \label{eq:3-8}
\end{equation}
In the last equality we have used (\ref{eq:3-5}). 
The last expression in (\ref{eq:3-8}) is useful for numerical calculations when $\Omega ^c$ is smaller than $\Omega $, 
because the indices of $A$ and $D$ take over only the space points on $\Omega ^c$ and the matrix sizes of $A$ and $D$ are smaller than those of $B$ and $E$. 
It can be shown that all of $\lambda_n$ are non-negative as follows.
From (\ref{eq:3-5}) we have
\begin{equation}
A \Lambda = -AEB^T = BFB^T .    \label{eq:3-9-1}
\end{equation}
It is easy to show that $A,C,D$ and $F$ are positive definite matrices when $W$ and $W^{-1}$ are positive definite matrices.
Then $A\Lambda$ is a positive semidefinite matrix as can be seen from (\ref{eq:3-9-1}).
So all eigenvalues of $\Lambda$ are non-negative.
After all, we can obtain the entanglement entropy by solving the eigenvalue problem of $\Lambda $.


\section{lattice formulation} \label {lattice}
We apply the above formalism to a massless free scalar field in $(d+1)$-dimensional Minkowski spacetime.
The Lagrangian is given by 
\begin{equation}
L= \int d^dx \dfrac{1}{2} [ \dot\phi ^2 - ( \nabla \phi  )^2 ].  \label{eq:4-1}
\end{equation}
As an ultraviolet regulator, we replace the continuous $d$-dimensional space coordinates $x$ by a lattice of discrete points with spacing $a$. 
As an infrared cutoff, we allow the individual components of $n \equiv x/a$ to assume only a finite number $N$ of independent values $-N/2<n_{\mu} \leq N/2 .$
The Greek indices denoting vector quantities run from one to $d$.
Outside this range we assume the lattice is periodic.
The dimensionless Hamiltonian $H_0\equiv a H$ is given by
\begin{equation}
H_0\equiv a H=  \sum_{n} [\dfrac{1}{2} \pi_{n}^2 +\dfrac{1}{2} \sum_{\mu =1}^d (\phi _{n_\nu +\delta _{\nu \mu }} -\phi _{n_\nu } )^2 +\dfrac{a^2 m^2}{2} \phi_{n}^2]
 \equiv \sum_{n} \dfrac{1}{2} \pi_{n}^2 + \sum_{m,n} \dfrac{1}{2} \phi_{m} V_{mn} \phi_{n} , \label{eq:4-2}
\end{equation}
where $\phi_{n}$ and $\pi_{n}$ are dimensionless and Hermitian, and obey the canonical commutation relations 
\begin{equation}
[\phi_{n} , \pi_{m}]=i \delta_{nm}  . \label{eq:4-3}
\end{equation}
In Eq.(\ref{eq:4-2}) we insert a mass term in order to remove a zero eigenvalue of $V_{mn}$; 
if $V_{mn}$ should have the zero eigenvalue, $W^{-1}$ in (\ref{eq:3-4}) would not exist.
Later we will take $N$ to infinity.
In this limit we can neglect the zero eigenvalue of $V_{mn}$ and will take $a m$ to zero. 
Taking $N$ to infinity is important in order to calculate the entanglement entropy $S_{AB}$ of two spheres. 
The entanglement entropy of two spheres is more sensitive to the value of $N$ than that of one sphere. 
(In fact, we numerically calculated $S_{AB}$ for finite $N$ with antiperiodic boundary conditions without the mass term. 
$S_{AB}$ depends on $N$ when the distance $r$ between two spheres is close to $N/2$, 
and we could not obtain the clear $r$ dependence of $S_{AB}$.) 

From (\ref{eq:4-2}) we obtain (see e.g. \cite{creutz1985quarks})
\begin{equation}
W_{m n}= N^{-d} \sum_{k} [a^2 m^2+2 \sum_{\mu =1}^{d} (1-\cos \dfrac{ 2\pi k_{\mu} }{N} ) ]^{1/2} e^{2\pi i k(n-m)/N} , \label{eq:4-4}
\end{equation}
\begin{equation}
W_{m n}^{-1}= N^{-d} \sum_{k} [a^2 m^2+2 \sum_{\mu =1}^{d} (1-\cos \dfrac{ 2\pi k_{\mu} }{N} ) ]^{-1/2} e^{2\pi i k(n-m)/N} , \label{eq:4-5}
\end{equation}
where the index $k$ also carries $d$ integer valued components, each in the range of $-N/2<k_{\mu} \leq N/2 $.
We take $N$ to infinity and change the momentum sum into an integral with the replacements $q_{\mu}=2\pi k_{\mu} /N$ and $N^{-d} \sum_{k}\rightarrow \int_{-\pi}^{\pi} \tfrac{d^d q}{(2\pi )^d}$,
and then we have
\begin{equation}
W_{m n}= \int_{-\pi}^{\pi} \dfrac{d^d q}{(2\pi )^d} e^{i q (n-m)} [a^2 m^2 + 2 \sum_{\mu =1}^{d} (1-\cos q_{\mu} ) ]^{\tfrac{1}{2}}  , \label{eq:4-6}
\end{equation}
\begin{equation}
W_{m n}^{-1}= \int_{-\pi}^{\pi} \dfrac{d^d q}{(2\pi )^d} e^{i q (n-m)} [a^2 m^2 + 2 \sum_{\mu =1}^{d} (1-\cos q_{\mu} ) ]^{\tfrac{-1}{2}}  . \label{eq:4-7}
\end{equation}
In (\ref{eq:4-6}) and (\ref{eq:4-7}) the integrals converge when $a m\rightarrow 0$, so we can take $a m$ to zero,
\begin{equation}
W_{m n}= \int_{-\pi}^{\pi} \dfrac{d^d q}{(2\pi )^d} e^{i q (n-m)} [2 \sum_{\mu =1}^{d} (1-\cos q_{\mu} ) ]^{\tfrac{1}{2}}  , \label{eq:4-8}
\end{equation}
\begin{equation}
W_{m n}^{-1}= \int_{-\pi}^{\pi} \dfrac{d^d q}{(2\pi )^d} e^{i q (n-m)} [ 2 \sum_{\mu =1}^{d} (1-\cos q_{\mu} ) ]^{\tfrac{-1}{2}}  . \label{eq:4-9}
\end{equation}

From (\ref{eq:4-8}) and (\ref{eq:4-9}) we can compute $W_{m n}$ and $W_{m n}^{-1}$ numerically.
Then we can compute the entanglement entropy from (\ref{eq:new3-1}), (\ref{eq:3-7}) and (\ref{eq:3-8}).
The integrands in (\ref{eq:4-8}) and (\ref{eq:4-9}) highly oscillate when $\| n-m\| \gg 1$, and the 
numerical integrals converge very slowly.
We can obtain approximate expressions of $W_{m n}$ and $W_{m n}^{-1}$ by hand when $\| n-m\| \gg 1$, 
so we will use them when $\| n-m\| \gg 1$ in order to reduce the computational complexity of $W_{m n}$ and $W^{-1}_{m n}$.
To evaluate $W_{m n}$ and $W_{m n}^{-1}$ when $\| n-m\| \gg 1$, we define $r\equiv a (n-m)$ and take $\| n-m\| $ to infinity 
keeping $r$ fixed.
We change the variable as $p=q/a$, and then we have
\begin{equation}
W_{m n}= a^d \int_{-\tfrac{\pi}{a} }^{ \tfrac{\pi}{a} } \dfrac{d^d p}{(2\pi )^d} e^{i p r} [ 2 \sum_{\mu =1}^{d} (1-\cos a p_{\mu} ) ]^{\tfrac{1}{2}} 
\rightarrow a^{d+1} \int_{-\infty }^{\infty } \dfrac{d^d p}{(2\pi )^d} e^{i p r -\tfrac{a}{\pi} \| p \|  } [\| p \|^2 ]^{\tfrac{1}{2}}  . \label{eq:4-10}
\end{equation}
We can perform the integral in (\ref{eq:4-10}) analytically when $\| r\| /a \rightarrow \infty $ (see Appendix A of \cite{Shiba:2010dy} ), and then we obtain
\begin{equation}
W_{m n} \rightarrow a^{d+1} \dfrac{A_d}{  \| r \|^{d+1} } =\dfrac{A_d}{  \| n-m \|^{d+1} }  ,   \label{eq:4-11}
\end{equation}
where   
\begin{equation}
A_d = \begin{cases} 
               -\dfrac{(d-1)!!}{(2\pi )^{d/2}}  & \textrm{for even} ~~ d\geq 2  , \\
                -2\dfrac{(d-1)!!}{(2\pi )^{(d+1)/2}}  & \textrm{for odd} ~~ d\geq 3  . \\   
              \end{cases}          \label{eq:4-12}
\end{equation}
We can evaluate $W_{m n}^{-1}$ when $\| n-m\| \gg 1$ in the same way (see Appendix A of \cite{Shiba:2010dy} ), and then we obtain
\begin{equation}
W_{m n}^{-1} \rightarrow a^{d-1} \int_{-\infty }^{\infty } \dfrac{d^d p}{(2\pi )^d} e^{i p r -\tfrac{a}{\pi} \| \vec{p} \|  } [\| \vec{p} \|^2 ]^{\tfrac{-1}{2}} 
\rightarrow  a^{d-1} \dfrac{B_d}{  \| r \|^{d-1} } =\dfrac{B_d}{  \| n-m \|^{d-1} }    . \label{eq:4-13}
\end{equation}
where 
\begin{equation}
B_d = \begin{cases} 
               \dfrac{(d-3)!!}{(2\pi )^{d/2}}  & \textrm{for even} ~~ d\geq 2  , \\
                2\dfrac{(d-3)!!}{(2\pi )^{(d+1)/2}}  & \textrm{for odd} ~~ d\geq 3  , \\   
              \end{cases}          \label{eq:4-14}
\end{equation}
where $0!!=(-1)!!=1.$


\section{numerical calculations} \label {results}
We calculate numerically the entanglement entropy $S_{AB}$ of two spheres $A$ and $B$ whose radii are $R_1$ and $R_2$, 
and the distance between the centers of them is $r$ for $d=3$.

We put the centers of the spheres on a lattice. 
We define the sphere whose radius is $R$ as a set of points which are at distances of  $R$ or less
 from the center of the sphere.
In order to reduce the computational complexity of $W_{m n}$ and $W^{-1}_{m n}$, 
we use the approximate expressions (\ref{eq:4-11}) and (\ref{eq:4-13}) when $\| n-m\| > 10$,
and we use the numerical integrals of exact expressions (\ref{eq:4-8}) and (\ref{eq:4-9}) when  $\| n-m\| \leq 10$.
When $\| n-m\| = 10$, the differences between the numerical integrals of the exact expressions and 
the approximate expressions are less than 4\% for $W_{m n}$ and less than 1\% for $W^{-1}_{m n}$. 
We perform matrix operations and calculate the eigenvalues $\lambda_n$ of the matrix $\Lambda $ in (\ref{eq:3-8}) with \texttt{Mathematica 8}.
The number of columns and rows of $\Lambda $ is the number of points in the region of which we calculate entanglement entropy.

\begin{figure}
 \includegraphics[width=8cm,clip]{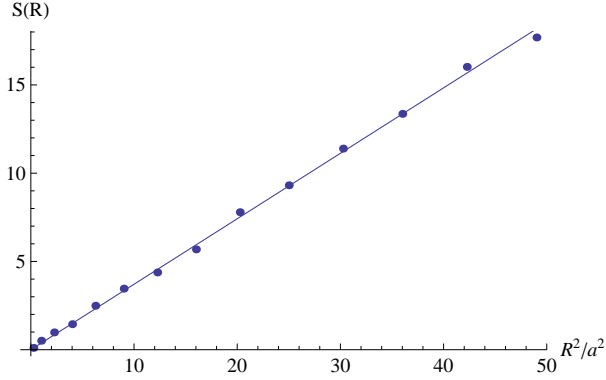}%
 \caption{The entanglement entropy $S(R)$ of one sphere whose radius is $R$
as a fumction of $R^2/a^2$. The line is the best linear fit.}
 \label{S(R)}
 \end{figure}

We show the computed values of $S(R)$ which is the entanglement entropy of one sphere 
as a fumction of $R^2/a^2$ in Fig.\ref{S(R)}, where $a$ is a lattice spacing.
The points are fitted by a straight line: 
\begin{equation}
S=0.37R^2/a^2.    \label{eq:5-1-1} 
\end{equation}
This result agrees with the result in \cite{Srednicki:1993im} except for the coefficient. 
(The coefficient in \cite{Srednicki:1993im} is 0.30.
This difference necessarily arises from the difference of regularization methods.
In \cite{Srednicki:1993im} the author use the polar coordinate system and replace the continuous radial coordinate by a lattice.)

\begin{figure}
 \includegraphics[width=8cm,clip]{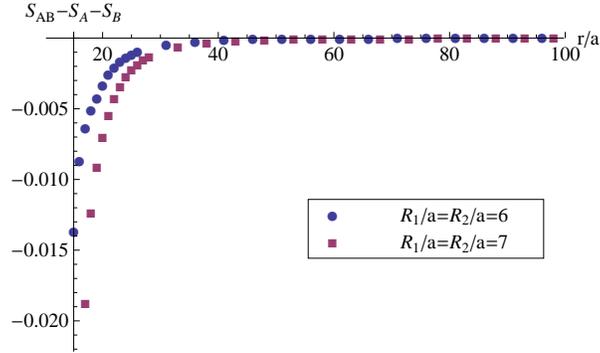}%
 \caption{ $S_{AB} -S_A-S_B$ as a function of $r/a$ for $R_1/a=R_2/a=6,7$, where $S_{AB}$ is the entanglement entropy of two spheres 
$A$ and $B$ whose radii are $R_1$ and $R_2$. 
The distance between the centers of the two spheres is $r$. }
 \label{SAB}
 \end{figure}

\begin{figure}
 \includegraphics[width=8cm,clip]{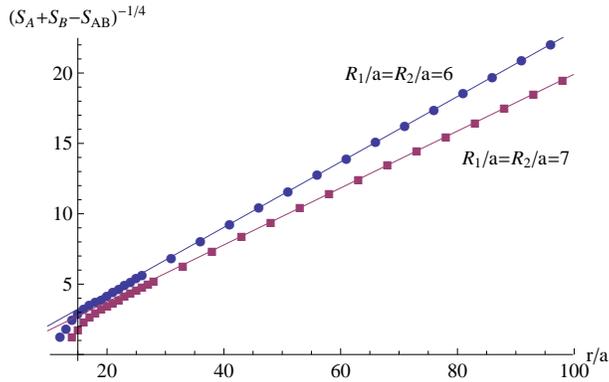}%
 \caption{ $(S_{A} +S_B-S_{AB})^{-1/4}$ as a function of $r/a$ for $R_1/a=R_2/a=6,7$, where $S_{AB}$ is the entanglement entropy of two spheres 
$A$ and $B$ whose radii are $R_1$ and $R_2$. 
The distance between the centers of the two spheres is $r$. 
The straight lines are fitted by the data between $r/a=R_1/a+R_2/a+24$ and  $r/a=R_1/a+R_2/a+84$. 
For $r\gtrapprox 3R(\equiv R_1=R_2) $ the lines are beautifully fitted and the approximate expressions (\ref{eq:5-1}) and (\ref{eq:5-2}) are precise. }
 \label{SAB14}
 \end{figure}

\begin{figure}
 \includegraphics[width=8cm,clip]{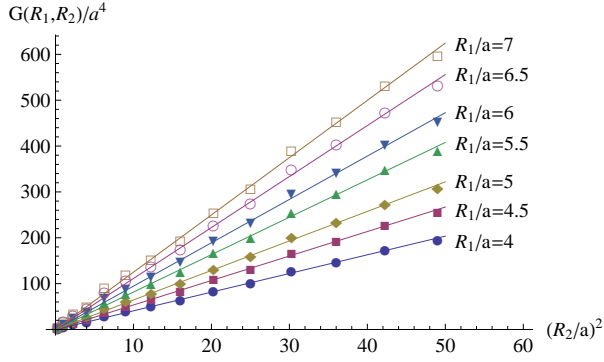}%
 \caption{$G(R_1,R_2)/a^4$ in (\ref{eq:5-1})
as a function of $R_{2}^{2}/a^2$ for $R_1/a=4,4.5,\dots,7$. The lines are the best linear fit.}
 \label{G}
 \end{figure}

\begin{figure}
 \includegraphics[width=8cm,clip]{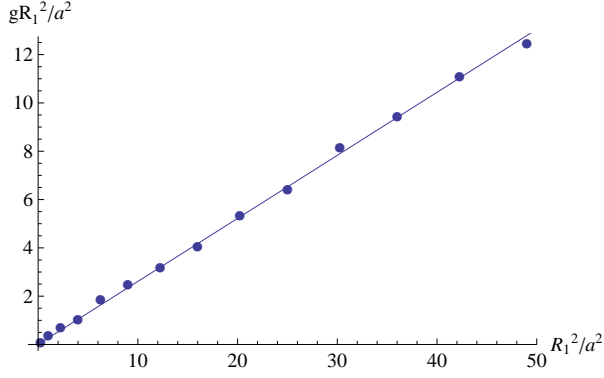}%
 \caption{$g R_1^2/a^2$ as a function of $R_1^2/a^2$, where $g$ is defined as $G(R_1,R_2)=g R_1^2 R_2^2$. The line is the best linear fit.}
 \label{gR2}
 \end{figure}

We show the computed values of $S_{AB}(r,R_1,R_2)$  which is the entanglement entropy of two spheres 
as a function of $r/a$ for $R_1/a=R_2/a=6,7$ in Fig.\ref{SAB}.
As can be seen, $S_{AB}$ reaches its maximum value $S_A+S_B$ when $r \rightarrow \infty$.
In order to clarify the behavior of $S_{AB}$ as a function of $r$, 
we show the computed values of $(S_A+S_B-S_{AB})^{-1/4}(r,R_1,R_2)$ 
as a function of $r/a$ for $R_1/a=R_2/a=6,7$ in Fig.\ref{SAB14}.
The straight lines in Fig.\ref{SAB14} are fitted by the data between $r/a=R_1/a+R_2/a+24$ and  $r/a=R_1/a+R_2/a+84$.
In these regions the points are beautifully fitted by the straight lines.
Then, when $r\gg R_1,R_2$, we obtain
\begin{equation}
-S_{A;B}\equiv S_{AB}(r,R_1,R_2) -S_A(R_1)-S_B(R_2) \approx -\dfrac{G(R_1,R_2)}{r^4} ,     \label{eq:5-1}
\end{equation}
where $G(R_1,R_2)$ is defined in (\ref{eq:5-1}) and $G(R_1,R_2)=G(R_2,R_1)\geq 0.$  
$S_{A;B}$ is the mutual information of $A$ and $B$. 
From Fig.\ref{SAB14} the approximate expression (\ref{eq:5-1}) is precise for relatively small $r$. 
(When $R_1=R_2\equiv R$, for $r\gtrapprox 3R$ (\ref{eq:5-1}) is precise from Fig.\ref{SAB14}.) 
We can obtain $G(R_1,R_2)/a^4$ from slopes of graphs of $(S_A+S_B-S_{AB})^{-1/4}(r,R_1,R_2)$, 
and then we show the computed values of $G(R_1,R_2)/a^4$ 
as a function of $R_{2}^{2}/a^2$ for $R_1/a=4,4.5,\dots,7$ in Fig.\ref{G}.
From Fig.\ref{G} we can see that $G(R_1,R_2)/a^4$ is proportional to $R_{2}^2$.
Because $G(R_1,R_2)=G(R_2,R_1)$, we obtain $G(R_1,R_2)=g R_1^2 R_2^2$, where $g$ is a dimensionless constant.
We can obtain the values of $g R_1^2$ from slopes of graphs of $G(R_1,R_2)$ as a function of $R_2^2$. 
To obtain the precise value of $g$, we show the computed values of $g R_1^2/a^2$ as a function of $R_1^2/a^2$ in Fig.\ref{gR2} 
and obtain $g=0.26$ from the slope of the line which is the best linear fit in Fig.\ref{gR2}.  

Finally, when  $r\gg R_1,R_2$, we obtain
\begin{equation}
-S_{A;B}= S_{AB}(r,R_1,R_2) -S_A(R_1)-S_B(R_2) \approx -\dfrac{0.26 R_1^2 R_2^2}{r^4} .     \label{eq:5-2}
\end{equation}
When $r\approx R_1,R_2$, from Fig.\ref{SAB14}, $S_{AB}$ rapidly decreases when $r$ decreases. 
(Note that we cannot determine the functional form of $G(R_1,R_2)$ only by the constraints from dimensional analysis, symmetry, and behavior in the limit $R_1\rightarrow 0$.    
For example, $G(R_1,R_2)=R_1^3 R_2+R_1 R_2^3$ is not prohibited by these constraints.)   

\begin{figure}
 \includegraphics[width=8cm,clip]{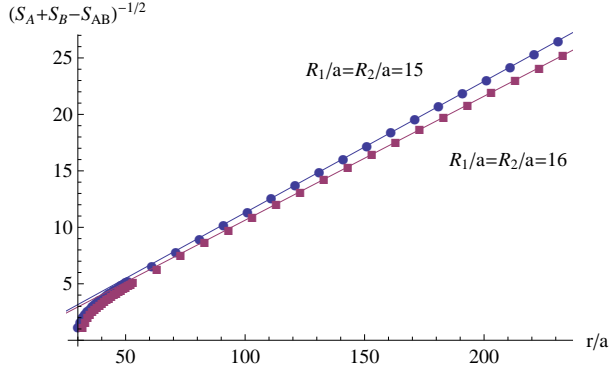}%
 \caption{ $(S_{A} +S_B-S_{AB})^{-1/2}$ as a function of $r/a$ for $R_1/a=R_2/a=15,16$, where $S_{AB}$ is the entanglement entropy of two discs 
$A$ and $B$ whose radii are $R_1$ and $R_2$. 
The distance between the centers of the two discs is $r$. 
The straight lines are fitted by the data between $r/a=R_1/a+R_2/a+101$ and  $r/a=R_1/a+R_2/a+201$. 
For $r\gtrapprox 4R(\equiv R_1=R_2) $ the lines are beautifully fitted and the approximate expression (\ref{eq:5-3}) is precise. }
 \label{d2}
 \end{figure}

For $d=2$, we compute $S_{AB}$ in the same way. 
We show only the computed values of $(S_A+S_B-S_{AB})^{-1/2}(r,R_1,R_2)$ 
as a function of $r/a$ for $R_1/a=R_2/a=15,16$ in Fig.\ref{d2}.
The straight lines in Fig.\ref{d2} are fitted by the data between $r/a=R_1/a+R_2/a+101$ and  $r/a=R_1/a+R_2/a+201$.
In these regions the points are beautifully fitted by the straight lines. 
We cross-checked our numerical procedure with the data of related calculations in Figure 1 in \cite{Casini:2009sr}.
(In the figure, the authors show the mutual information of two discs for $R_1=R_2=R$ and $r=3R$. 
Our results were very close to theirs.)
Finally, when $r\gg R_1,R_2$, we obtain
\begin{equation}
-S_{A;B}= S_{AB}(r,R_1,R_2) -S_A(R_1)-S_B(R_2) \approx -\dfrac{0.37 R_1 R_2}{r^2} .     \label{eq:5-3}
\end{equation}

\begin{figure}
 \includegraphics[width=8cm,clip]{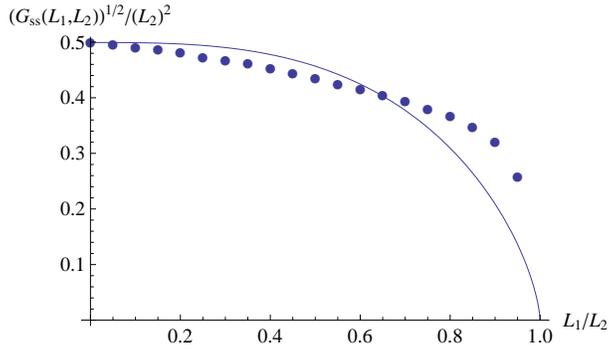}%
 \caption{$(G_{ss}(L_1,L_2))^{1/2}/L_2^2$ for $L_2=10a$ as a function of $L_1/L_2$. 
$G_{ss}(L_1,L_2)$ is defined as $S_{D;E} \approx G_{ss}(L_1,L_2)/r^{4}$ when $r\gg L_1,L_2$. 
The curve is $0.50(1-(L_1/L_2)^{3})^{2/3}$. 
$(G_{ss}(L_1,L_2))^{1/2}/L_2^2$ is monotone decreasing function of $L_1/L_2$ and not proportional to $(1-(L_1/L_2)^{3})^{2/3}$.}
 \label{Gssd3}
 \end{figure}

\begin{figure}
 \includegraphics[width=8cm,clip]{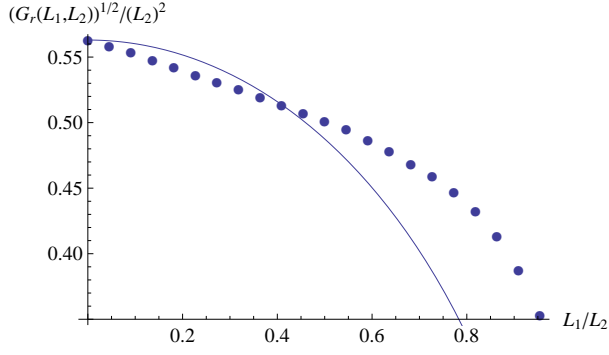}%
 \caption{$(G_{r}(L_1,L_2))^{1/2}/L_2^2$ for $L_2=22a$ as a function of $L_1/L_2$. 
$G_{r}(L_1,L_2)$ is defined as $S_{H;I} \approx G_{r}(L_1,L_2)/r^{2}$ when $r\gg L_1,L_2$. 
The curve is $0.56(1-(L_1/L_2)^{2})^{1/2}$.
$(G_{r}(L_1,L_2))^{1/2}/L_2^2$ is monotone decreasing function of $L_1/L_2$ and not proportional to $(1-(L_1/L_2)^{2})^{1/2}$. }
 \label{Grd2}
 \end{figure}

In order to examine whether only the degrees of freedom on the surface of the spheres contribute to the mutual information or not, 
we calculate the mutual information $S_{D;E}$ of two same spherical shells $D$ and $E$ for $d=3$ and 
the mutual information $S_{H;I}$ of two same rings $H$ and $I$ for $d=2$. 
The internal (external) radii of the spherical shell and the ring are $L_1$ ($L_2$). 
The distance between the centers of the two spherical shells and that between the two rings are $r$.
When $r\gg L_2$, we obtain $S_{D;E} \approx G_{ss}(L_1,L_2)/r^{4}$ and $S_{H;I} \approx G_{r}(L_1,L_2)/r^{2}$. 
We show $(G_{ss}(L_1,L_2))^{1/2}/L_2^2$ for $L_2=10a$ as a function of $L_1/L_2$ in Fig.\ref{Gssd3}
and $(G_{r}(L_1,L_2))^{1/2}/L_2$ for $L_2=22a$ as a function of $L_1/L_2$ in Fig.\ref{Grd2}. 
The curve in Fig.\ref{Gssd3} is $0.50(1-(L_1/L_2)^3)^{2/3}$ and 
the curve in Fig.\ref{Grd2} is $0.56(1-(L_1/L_2)^2)^{1/2}$. 
We show these curves for comparison with the data.  
From Fig.\ref{Gssd3} and Fig.\ref{Grd2}, $(G_{ss}(L_1,L_2))^{1/2}/L_2^2$ and $(G_{r}(L_1,L_2))^{1/2}/L_2$ are monotone decreasing function of $L_1/L_2$,  
and $(G_{ss}(L_1,L_2))^{1/2}$ is not proportional to the $2/3$ power of the volume of the spherical shell, 
and $(G_{r}(L_1,L_2))^{1/2}$ is not proportional to the $1/2$ power of the area of the ring. 
Then not only the degrees of freedom on the surface of the sphere but also those on the inside region contribute to the mutual information,   
and the degrees of freedom on the inside region does not contribute uniformly to the mutual information. 

\begin{figure}
 \includegraphics[width=8cm,clip]{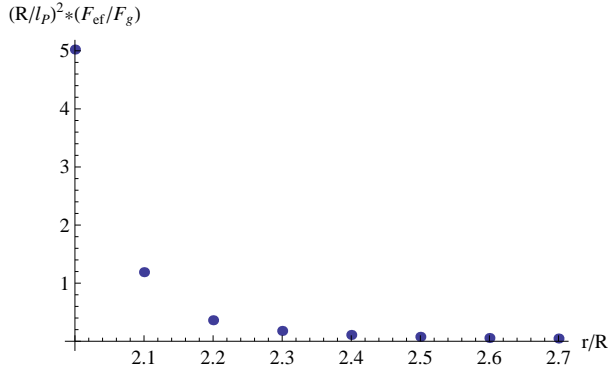}%
 \caption{The ratio of the entropic force to the force of gravity $\left( \tfrac{R}{l_P} \right)^2  \left| \tfrac{F_{ef}}{F_g}\right| $  
as a function of $r/R$ for $R/a=10$. }
 \label{Fef}
 \end{figure}

We roughly estimate the magnitude of the entropic force between two black holes by using $S_{AB}$ 
in Minkowski spacetime.
We consider two black holes ($A$ and $B$) which have the same radius $R_1=R_2\equiv R$ and the distance between which is $r$.  
For simplicity, we consider the case that the state of the field on the whole space is a pure state. 
Generally, if a composite system $XY$ is in a pure state, then $S_X=S_Y$ \cite{nielsen2000quantum}.
Then the entanglement entropy of the outside region of two black holes is equal to that of 
the inside regions of two black holes.
We define the entropic force of the field on the outside region which acts on one black hole in 
the direction of increasing $r$ as $F_{ef}$. 
$F_{ef}$ is given by
\begin{equation}
F_{ef}=T\dfrac{\partial S_{AB}}{\partial r},     \label{eq:5-4}
\end{equation}
where $T$ is the temperature of the field of the outside region. 
To estimate $F_{ef}$, we set $S_{AB}$ to that in Minkowski spacetime 
and $T$ to the Hawking temperature $T=(8\pi G_{N} M)^{-1} =(4\pi R)^{-1}$.
In this approximation the entropic force is \textit{repulsion force} because $S_{AB}$ increases when $r$ increases.
$\partial S_{AB}/ \partial r$ is independent of the ultraviolet cutoff, 
and then we obtain
\begin{equation}
F_{ef}=-\dfrac{T}{R} S'_{A;B}(r/R)=-\dfrac{1}{4\pi R^2} S'_{A;B}(r/R),     \label{eq:5-5}
\end{equation}
where $S_{A;B}=S_A(R)+S_B(R)-S_{AB}(r,R)$ and $S'_{A;B}\equiv \partial S_{A;B}/ \partial (r/R)$. 
($S_{A;B}$ is independent of the ultraviolet cutoff and a function of $r/R$.) 
Then the ratio of the entropic force to the force of gravity ($F_g=-\tfrac{G_N M^2}{r^2}=-\tfrac{R^2}{4G_N r^2}$) is 
\begin{equation}
\left| \dfrac{F_{ef}}{F_g}\right|=\dfrac{1}{\pi} \left( \dfrac{l_P}{R} \right)^2 
\left| \dfrac{S'_{A;B}(r/R)}{R^2/r^2}\right|,     \label{eq:5-6}
\end{equation}
where $l_P$ is the Planck length $l_{P}=(G_N \hbar /c^3)^{1/2}$.
When $r\gg R$, we substitute (\ref{eq:5-2}) into (\ref{eq:5-6}), 
and then we obtain 
\begin{equation}
\left| \dfrac{F_{ef}}{F_g}\right| \approx 0.33 \left( \dfrac{l_P}{R} \right)^2 
\dfrac{R^3}{r^3}.     \label{eq:5-7}
\end{equation}
When $r\approx R$, we show the computed values of 
$\left( \tfrac{R}{l_P} \right)^2  \left| \tfrac{F_{ef}}{F_g}\right| $  
as a function of $r/R$ for $R/a=10$ in Fig.\ref{Fef}. 
(Although from (\ref{eq:5-6}) $\left( \tfrac{R}{l_P} \right)^2  \left| \tfrac{F_{ef}}{F_g}\right| $ is a function of $r/R$ 
and is independent of the choice of the value of $R/a$, the computed values of $\left( \tfrac{R}{l_P} \right)^2  \left| \tfrac{F_{ef}}{F_g}\right| $ 
slightly depend on the choice of the value of $R/a$ because the spheres on the lattice are distored. 
When $R/a$ is large, the spheres on the lattice are similar to the real spheres and this $R/a$ dependence is small.) 
From (\ref{eq:5-7}) and Fig.\ref{Fef} the entropic force is much smaller than the force of gravity when $R\gg l_P$ 
and comparable to the force of gravity when $R\approx l_P$. 


\section{conclusions and discussion} \label {conclusion}
We calculated numerically the entanglement entropy $S_{AB}$ of two spheres 
and obtained the approximate expression (\ref{eq:5-2}). 
From Fig.\ref{SAB14}, (\ref{eq:5-2}) is precise for relatively small $r$. 
(When $R_1=R_2\equiv R$, for $r\gtrapprox 3R$ (\ref{eq:5-2}) is precise from Fig.\ref{SAB14}.) 
We showed that the mutual information $S_{A;B}$ of $A$ and $B$ is independent of 
the ultraviolet cutoff for $d=2,3$ though $S_A$ and $S_B$ depends on the ultraviolet cutoff. 
The mutual information $S_{A;B}$ measures the entanglement between $A$ and $B$ 
and $S_A$ measures the entanglement between $A$ and $A^c$ where $A^c$ is the complementary of $A$.
Then our results mean that the ultraviolet divergence of entanglement entropy in QFT is caused by the 
entanglement between points which are infinitely close to each other and the entanglement between 
regions which are finitely separate from each other is finite. 
And we showed that $S_{A;B}$ is the simple product of a function of $R_1$ and that of $R_2$ for $d=2,3$. 
These properties of $S_{A;B}$ for $d=2,3$ are most likely the same as those for $d\geq 4$. 
Then, from (\ref{eq:1-1}), for $d\geq 4$ when $r\gg R_1,R_2$ we assume 
\begin{equation}
S_{A;B}\approx \dfrac{g_d R_1^{d-1} R_2^{d-1}}{r^{2d-2}}, \label{eq:6-1}
\end{equation}  
where $g_d\geq 0$ is a dimensionless constant.

In order to examine whether only the degrees of freedom on the surface of the spheres contribute to the mutual information or not, 
we calculate the mutual information $S_{D;E}$ of two same spherical shells $D$ and $E$ for $d=3$ and 
the mutual information $S_{H;I}$ of two same rings $H$ and $I$ for $d=2$. 
We obtained the result that 
not only the degrees of freedom on the surface of the sphere but also those on the inside region contribute to the mutual information,    
and the degrees of freedom on the inside region does not contribute uniformly to the mutual information. 
Because $S_{D;E}$ and $S_{H;I}$ measure the entanglement between regions which are finitely separate from each other, 
it is natural that the inside region contribute to the mutual information. 
The result that the inside region does not contribute uniformly to the mutual information means that 
the mutual information is not the product of the simple sum of the contribution from each volume elements. 
These results are different from that of the entanglement entropy to which the degrees of freedom on the surface of the boundary contribute mainly and uniformly. 
So the mutual information of two disconnected regions is not universally proportional to the product of the surface areas of the regions. 
Because a sphere has only one dimensionful parameter, the mutual information of two spheres is proportional to the product of the surface areas.
For example, the mutual information of two rectangular solids is most likely not proportional to the product of the surface areas 
because a rectangular solid has three dimensionful parameters.  



Our numerical method has three properties. 
First, 
we take the volume of the whole space to infinity, i.e. $N\rightarrow \infty $ in (\ref{eq:4-4}) and (\ref{eq:4-5}).
Second, the computational complexity of our method depends only on the number of points on the regions of which we compute the entanglement entropy 
and does not depend on the distance between the separated regions. 
The computational complexity of conventional methods increases when the distance between the separated regions increases.  
This is because the numerical integrals of $W_{m n}$ in (\ref{eq:4-8}) and $W^{-1}_{m n}$ in (\ref{eq:4-9}) converge very slowly when $\| n-m\| \gg 1$.
In order to reduce the computational complexity of $W_{m n}$ and $W^{-1}_{m n}$, 
we use the approximate expressions (\ref{eq:4-11}) and (\ref{eq:4-13})  when $\| n-m\| > 10$. 
Third, we can compute the entanglement entropy of general shaped regions by our method 
because we do not use any symmetry of the regions of which we compute the entanglement entropy in our method. 
For example, we can compute the entanglement entropy of more than two separated regions.
The first and the second properties enable us to obtain the $r$ dependence of $S_{AB}$. 
And the third property enable us to compute $S_{AB}$ for $R_1\neq R_2$. 

We estimated roughly the magnitude of the entropic force between two black holes. 
From (\ref{eq:5-7}) and Fig.\ref{Fef} the entropic force is comparable to the force of gravity when $R\approx l_P$. 
This rough estimate suggests that the entropic force is important for Planck scale black holes. 
(Of course, this result would be changed if the effect of quantum gravity would be taken into account when $R\approx l_P$.)

Next, we discuss the microscopic origin of the entropic force. 
As we see from (\ref{eq:5-5}) the entropic force is proportional to the $r$ derivative of the mutual information $S_{A;B}$.  
So the origin of the entropic force is the entanglement between inside regions of two black holes. 
Due to the entanglement between inside regions of two black holes, the density matrix of the scalar field on the outside region changes when $r$ changes. 
Then the force acts on black holes along the direction  in which $S_{AB}$ increases. 

Finally we mention the validity of this estimate.
When $r\gg R$, it is shown that $S_{A;B}$ in the black holes case can be expected to be similar to 
that in the Minkowski spacetime case except for the coefficient 
because almost all regions between two black holes is similar to Minkowski spacetime \cite{Shiba:2010dy}. 
So, the rough estimate corresponds to the contribution to $F_{ef}$ in (\ref{eq:5-4}) from $S_{A;B}$.
However, in the black holes case $S_{A}$ and $S_{B}$ depend on $r$ and contribute to $F_{ef}$. 
These contribution from $S_{A}$ and $S_{B}$ has been discussed in \cite{Shiba:2010dy}.
When $r \approx R$, $S_{A;B}$ in the black holes case is probably different from  
that in the Minkowski spacetime case 
because the region between two black holes is very different from Minkowski spacetime. 
However, even when $r \approx R$, $S_{A;B}$ is most likely independent of the ultraviolet cutoff as that in the Minkowski spacetime case. 

\begin{acknowledgments}
I am grateful to Takahiro Kubota and Satoshi Yamaguchi for a careful reading of
this manuscript and useful comments and discussions.
I also would like to thank Horacio Casini for informing me that I can cross-check my numerical procedure with the data 
of related calculations in \cite{Casini:2009sr}.
This work was supported
by a Grant-in-Aid from JSPS (No. 22-1930).
\end{acknowledgments}


\end{document}